# CROSS-CULTURAL MOOD PERCEPTION IN POP SONGS AND ITS ALIGNMENT WITH MOOD DETECTION ALGORITHMS

**Harin Lee**[1,2]   **Frank Höger**[2]   **Marc Schönwiesner**[1,3]   **Minsu Park**[4]   **Nori Jacoby**[2]

[1] Max Planck Institute for Human Cognitive and Brain Sciences
`hlee@cbs.mpg.de`
[2] Max Planck Institute for Empirical Aesthetics
`{frank.hoeger, nori.jacoby}@ae.mpg.de`
[3] Leipzig University
`marcs@uni-leipzig.de`
[4] New York University Abu Dhabi
`minsu.park@nyu.edu`

## ABSTRACT

Do people from different cultural backgrounds perceive the mood in music the same way? How closely do human ratings across different cultures approximate automatic mood detection algorithms that are often trained on corpora of predominantly Western popular music? Analyzing 166 participants' responses from Brazil, South Korea, and the US, we examined the similarity between the ratings of nine categories of perceived moods in music and estimated their alignment with four popular mood detection algorithms. We created a dataset of 360 recent pop songs drawn from major music charts of the countries and constructed semantically identical mood descriptors across English, Korean, and Portuguese languages. Multiple participants from the three countries rated their familiarity, preference, and perceived moods for a given song. Ratings were highly similar within and across cultures for basic mood attributes such as *sad*, *cheerful*, and *energetic*. However, we found significant cross-cultural differences for more complex characteristics such as *dreamy* and *love*. To our surprise, the results of mood detection algorithms were uniformly correlated across human ratings from all three countries and did not show a detectable bias towards any particular culture. Our study thus suggests that the mood detection algorithms can be considered as an objective measure at least within the popular music context.

## 1. INTRODUCTION

Music can express a range of emotions, from melancholy and sadness to love and joy. Since music has the powerful ability to reflect and modify one's emotional state [1], researchers have explored how people perceive emotion in music from diverse angles [2]. However, while the universality and variability of musical properties across cultures have been reported in many aspects [3,4], cross-cultural congruence of perceived moods in music has yet to reach a clear agreement [5]. In this paper, we present an empirical analysis of cross-cultural experiments that characterizes agreements and discrepancies on perceived moods in music by comparing between three cultures and also between human judgments and four algorithmic estimates.

### 1.1 Background

Many cognitive models have been proposed to account for emotional experience, and these can also be applied to account for the case of music. Basic emotion theory relies on discrete categories of emotions, whereas a competing explanation relies on dimensional models to argue that all affective states arise from a few independent affective dimensions [6]. In the context of music, previous studies have suggested that basic emotions such as happiness, sadness, and anger can be universally recognized in music by demonstrating that listeners can correctly deduce an intended emotion from unfamiliar musical traditions [7,8]. Nevertheless, critics have argued that the application of basic emotion theory to music is too simplistic and it fails to capture the full emotional richness expressed through music (see Eerola & Vuoskoski for a review [2]). Consequently, other scholars have proposed music-specific emotion models, including nine [9] or thirteen [10] dimensions.

Many researchers used such high-dimensional models and provided extensive empirical support. Yet, since such studies were based mainly on homogeneous Western participants (e.g., college students at a large research university [11]), it still remains unclear how these mood dimensions can be applied cross-culturally. Although significant cross-cultural differences were recently reported in multinational comparisons (e.g., [12,13]) or through non-US contexts (e.g., [14,15]), these studies are also limited in testing within a single musical style or a single population, respectively.

There is also a prevalent issue in the choice of lexical semantics when comparing between languages. That is, the subtle differences in the nuance of the mood terms translated into other languages can result in the varied interpretation of the meanings. Recent research has shown that the semantic alignment of emotion terms is highly variable across languages [16,17]. Thus, the cross-cultural differences observed in the aforementioned studies could have partly been driven by varied interpretations of word meanings rather than by the perception of music itself.

In the domain of music information retrieval (MIR), multiple automatic mood detection algorithms have been introduced within the last decade. These algorithms attempt to predict a listener's perception of high-level mood attributes in songs such as *danceability* and emotional *valence*. Such technology has shown to have diverse applications, including music recommendation [18], hit song prediction [19,20], and investigation of musical trends over time [21], all of which rely on quantifying large corpora of music. However, the generalizability of these tools is concerning because these mood detection algorithms are often trained on databases predominantly consisting of English-language songs judged by Western annotators (e.g., Billboard HOT100 [22] and Million Song Dataset [23]). It therefore raises important questions on how robustly these measures would perform on non-Western songs and reflect the perception of non-Western listeners.

### 1.2 Research Question

The challenges and opportunities in studying cross-cultural applicability of perceived moods in music lead us to introduce three research questions that guide the remainder of this paper:

**RQ1**. Do people with different cultural backgrounds perceive mood in music differently?
**RQ2**. Is there a more robust agreement within and across cultures for certain moods compared to others?
**RQ3**. How well do mood detection algorithms in MIR approximate human judgments, and is there a cultural bias in the algorithms?

We address the first two questions with a cross-cultural comparison between Brazilian, South Korean, and American raters, examining how they perceive nine types of moods in music. We chose these populations since they are geographically and culturally distant and speak different languages. Moreover, as recent evidence has shown that shared music interests around the world generate unique clusters in the West, Asia, and Latin America [1,24], we wanted to include one country from each cultural region in our study.

Many studies of musical emotions rely on film music excerpts or hand-picked musical stimuli that have been shown to evoke certain emotions [2]. In our design, we built a novel dataset of pop songs, randomly drawn from major music charts in those three countries, in order to test in a setting that is closer to real-world experience (see section 2.1). To begin, we used visual stimuli to validate whether the translated mood terms convey the exact intended meanings in the three languages (see section 3.3). We then ran identical online experiments in each country, asking participants to rate songs from all three cultures according to preference, familiarity, and nine semantic lexical mood descriptors (see section 2.3). We compared the similarity of mood ratings both within and across countries and identified a set of mood attributes that were more highly agreed upon than others.

To address RQ3, we used Spotify's API to retrieve four high-level mood features (*danceability*, *energy*, *valence*, and *acousticness*) and assessed how well these algorithms approximate human raters. Next, we examined whether there is a Western bias in the algorithms by testing if MIR values align better with raters from the US.

## 2. MATERIALS

### 2.1 Song Selection

We established a novel and balanced dataset of 360 pop songs originating from Brazil, S. Korea, and the US. We retrieved the major music charts for songs that made the charts between 2010 and 2019 (US: Billboard HOT 100[1], Brazil: Crowley Broadcast Analysis[2] and Top-40 Charts[3], Korea: Gaon Music Chart[4]).

We ran search queries for all unique songs using Spotify's Web API, retrieving a maximum of 50 results. For every search, a fuzzy string match ratio between 0 and 1 was computed based on the Levenshtein distance between the query and result strings, normalized by the maximum possible distance. We selected the entry that had the highest value and excluded songs with no results above a ratio of 0.7. We also manually inspected a subsample of 100 songs to validate the accuracy of the matching process (98% in Brazilian and US songs; 97% in Korean songs). Songs with the word *live* or *remix* in the title or without preview audio (to use as stimuli in the experiment) were also excluded.

The list was further reduced by retaining only the songs that were performed by artists matching the nationality of the chart (e.g., songs by Korean artists in the Korean charts) to balance the musical styles from the three countries. A song's origin was determined according to the International Standard Recording Code (ISRC), which begins with two alphabet letters that correspond to the standard ISO-2 country code to indicate the place of registration, and we manually inspected the final set.

We randomly sampled 12 songs per year in each country's list while controlling for duplicate artists. The final dataset consisted of 360 songs, that is, 120 songs from each country with songs distributed evenly across the 10-year window (the full list of songs and the experimental data are available at https://osf.io/3uw9d/).

### 2.2 MIR Mood Features

High-level audio features for all songs were retrieved using Spotify's Web API (see their reference manual for detailed descriptions of all available acoustic features[5]). Although other MIR libraries such as Essentia [25] also offer a similar set of features, it has been shown that audio codec and level of audio compression can influence the extraction outcome [26]. In fact, comparing Spotify's feature *danceability* with the same feature extracted using Essen-tia with low-quality preview audio resulted in small correlation ($r = .28$). Due to this limitation, we decided to only include Spotify's features in this study. Unfortunately, Spotify's algorithm is not open-source and we leave open the task of validating and comparing between different MIR algorithms to future researchers.

---

[1] Billboard Chart: https://www.billboard.com
[2] Crowley Broadcast Analysis: https://charts.crowley.com.br
[3] Top40 Charts: https://www.top40-charts.com
[4] Gaon Chart: http://www.gaonchart.co.kr
[5] Spotify Web API: https://developer.spotify.com/documentation/web-api

### 2.3 Selection of Mood Attributes

We used the nine-factor Geneva Emotion Music Scale (GEMS) [9] and the 13 emotion dimensions proposed by Cowen and colleagues [10] as a reference and chose the seven co-occurring dimensions in the two models as our target mood dimensions for investigation. These seven mood attributes were: *energetic/pump-up, relaxing/calm, dreamy, in love, joyful/cheerful, anxious/tense,* and *sad/depressing*. In addition, we added two extra mood attributes (*danceable* and *electronic*) that are directly comparable with the mood features retrieved through Spotify,[6] giving rise to nine mood attributes in total.

## 3. EXPERIMENT

### 3.1 Participants

We created independent participant pools in Brazil, S. Korea, and the US. The US participants were recruited from Amazon Mechanical Turk (MTurk) and had to pass a series of pre-screening tasks to qualify for the pool. The pre-screening tasks were: a LexTale task [27] to check for English fluency, a headphone screening task [28] to ensure they follow the instructions when asked to wear headphones, and an attention check item to screen for fraudulent behavior.

Since MTurk is not very popular in Brazil and is unavailable in S. Korea, participants from these countries were recruited independently by hiring research assistants in the local area. We required all participants to be native speakers and to reside in their respective countries.

We advertised identical experiments to all three participant pools with the content of the experiment having been translated into their respective languages (section 3.3 describes how we chose the most appropriate words in Korean and Portuguese for the nine mood attributes). In total, 166 participants participated in the study and 11,500 ratings were collected across the 360 songs (see Table 1).

|  | **Brazil** | **S. Korea** | **US** |
|---|---|---|---|
| **Participants (N)** | 58 (21 F) | 54 (27 F) | 54 (21 F) |
| **Ratings per mood** | 3,865 | 3,859 | 3,776 |
| **Average ratings per participant** | 66.6 ($SD$=33.1) | 71.5 ($SD$=33.1) | 69.9 ($SD$=33.1) |
| **Average ratings per song** | 10.7 ($SD$=2.74) | 10.7 ($SD$=2.31) | 10.5 ($SD$=3.02) |

**Table 1.** Collected ratings across 360 songs among the three countries. "F" denotes female participants.

### 3.2 Experiment Procedure

The stimuli set of 360 songs were divided into 12 blocks, with each block containing 10 songs from each country, evenly distributed across the years. This allowed participants to perform a varied number of blocks while always being presented with a counter-balanced set of stimuli across song origins and years.

After providing their informed consent, participants were randomly assigned to one of the 12 blocks, with each block containing 30 songs (see Figure 1). In each trial, they heard a 20 seconds snippet of a song randomly drawn from the block. They were then asked to identify the gender of the singer ("male", "female", or "don't know"), their familiarity of the song (4 choices ranging from "never heard of it" to "I know it very well"), preference (5-point scale from "dislike a lot" to "like a lot"), and their perception of nine moods presented in a random order (4-point scale from "not at all …" to "very …")[7]. Full questionnaire text is available in supplementary S2.

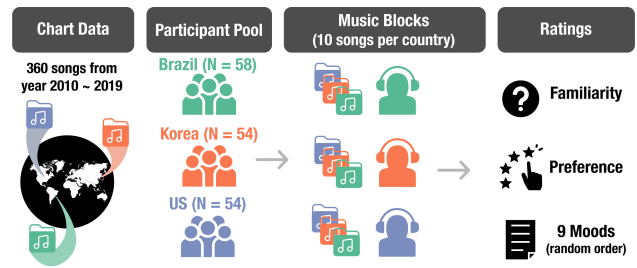

**Figure 1.** Experimental setup.

The experiment took around 40 minutes to complete and participants were compensated at an hourly rate of $9. Each participant could participate up to four times, with our system ensuring that they were never assigned to the same block more than once. The number of participants, gender ratio, and the average number of ratings per participant were well balanced across the three countries (see Table 1). All experiments were conducted using *PsyNet* [29], an automated recruitment framework for online experiments. Ethical approval was obtained by the Max Planck Society.

### 3.3 Validation of Mood Terms

Since the nine mood attributes were based in English, we had to translate those descriptors into Korean and Portuguese. However, word meanings can vary across cultures and this variability can lead to unintended biases [30]. Thus, to ensure that the intended meanings are interpreted in the same way across cultures with different languages, we validated the word meanings in a different modality (i.e., perception of images).

First, we preselected 17 Korean words and 20 Portuguese words that potentially align with the English reference words that we selected from the literature [9,10]. Second, we searched an online stock footage library[8] with the target mood terms and compiled a set of images (2 images per target mood term). We then recruited participants (38, 31, and 20 from the US, Brazil, and S. Korea, respectively) from the same participant pool recruited for the main experiment. We asked the participants to rate the relevance of a word to a displayed image on a 4-point scale. Finally, to find the best matching combinations of words across the languages, we computed Pearson correlations across the image items for all possible pairs of every mood dimension. We then selected the triplet word combinations with the highest overall correlations (see supplementary S1 for all word comparisons).

---

[6] Spotify's feature *acousticness* was paired with human judgment on *electronic* due to ambiguities in translations. Similarly, while *valence* from Spotify arises from a dimensional emotion model, it was paired with human judgment on *sad* as the closest matching discrete term.

[7] We also collected ratings on the most suitable color to the song. However, this is not discussed as it is beyond the scope of this paper.
[8] https://www.shutterstock.com

Figure 2B illustrates an example of how we found the best matching translations for mood *dreamy*. Using dictionary definitions for the word "dreamy" in Portuguese ("sonhador") and Korean ("몽환적인") yielded a mean correlation of $r = 0.73$, whereas the expression "like in a dream" in Korean ("꿈꾸는 듯한") and Portuguese ("como um sonho") resulted in considerably higher semantic alignment ($r = 0.88$). Thus, the latter combination was a more appropriate translation, and we used these words as the guiding passage in the main experiment for the mood dimension *dreamy*. The final combination of words chosen for the nine moods ranged in correlations between 0.88 and 0.98.

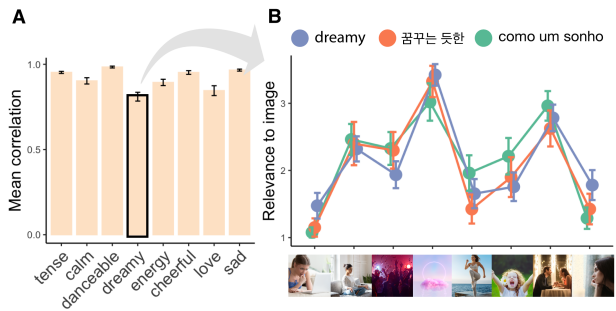

**Figure 2.** Matching mood semantics across English, Portuguese, and Korean with error bars representing 95% CI. (A) Correlations between word combinations across languages. (B) An example of best-matching triplet word combinations for *dreamy*.

## 4. RESULTS

### 4.1 Familiarity & Preference

Familiarity ratings provided by the three countries showed that the raters were most familiar with songs from their own country (see Figure 3A; all $p$s < .001 with Tukey's HSD), ensuring that our recruited participants and the songs in the dataset were adequate representative samples. While Brazilians and Koreans hardly knew each other's music (with a mean close to 1 = "never heard of it"), they were relatively familiar with American songs, reflecting the global presence of American pop songs. By contrast, raters from the US knew very little about both Korean and Brazilian songs.

Koreans and Americans also preferred their local music over foreign music (see Figure 3B; $p$s < .05 with Tukey's HSD), but the difference was small. Moreover, Brazilians did not prefer their local music over American music. This demonstrates that, while the raters may have been most familiar with their local music, they do not particularly prefer local pop songs over foreign ones.

However, when computing the correlations between familiarity and preference at individual participant level, there were strong correlations among all countries (Brazil: $r = .67$, Korea: $r = .66$, US: $r = .49$, all $p$s < .001; see Figure 3C). These results are consistent with numerous studies that have observed a strong relationship between a listener's familiarity and preference in music [31].

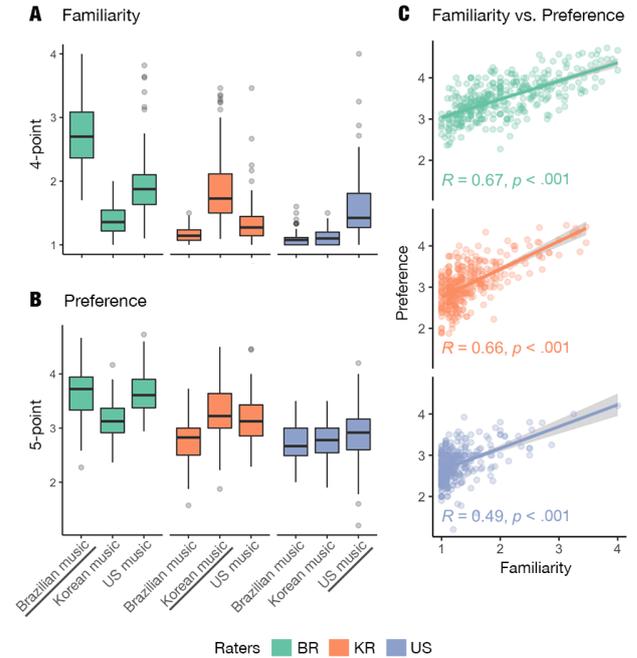

**Figure 3.** (A) Familiarity and (B) preference of songs among the three countries shown with box and whisker plot separated by song's origin. The underlined texts represent music that is local to the rater. (C) Scatter plot with general linear model fitting showing the relationship between familiarity and preference at the participant level.

### 4.2 Within Country Agreement

How are the ratings for the nine moods agreed upon among the raters within each country (RQ1)? Does agreement converge more for some moods than others (RQ2)?

To answer these questions, we examined rater agreement within each country by computing the intraclass correlation (ICC) using two-way random effects with the absolute agreement and multiple raters model [32]. Among the total of 12 blocks that divided the 360 song items, each block consisted on average 9.5 ($SD = 2.62$) unique raters from the same country who completed the same block. The ICC was computed for (i) each block separately and averaged across the 12 blocks for every mood variable (hereafter denoted as ICC$_{mean}$) and (ii) the perceived *gender of the singer* (used as baseline measure). As an alternative statistical measure, we also used split-half correlation to estimate the reliability. This showed nearly identical values and patterns to the ICC (for details, see supplementary S3).

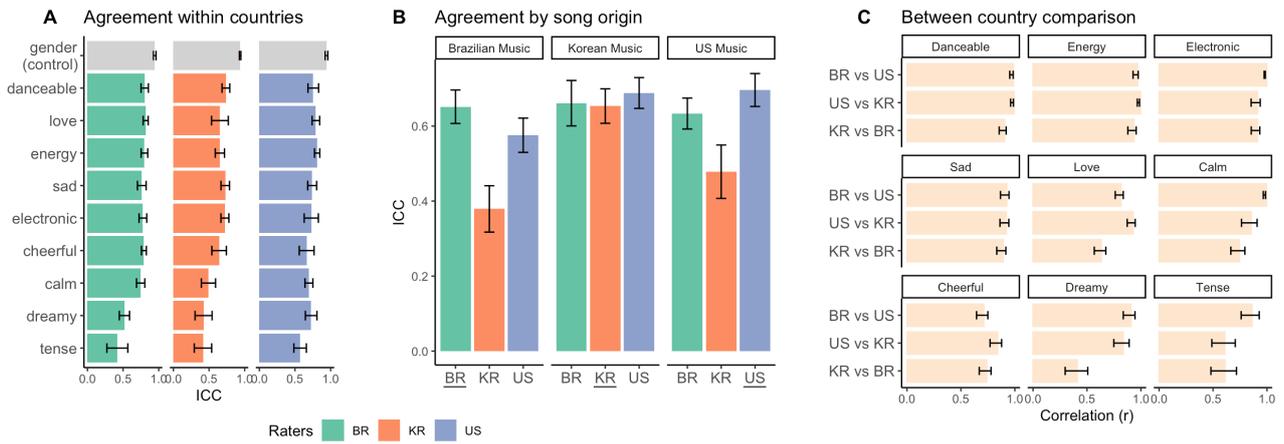

**Figure 4.** (A) Agreement across the moods within each country. The grey bars represent the baseline measure. (B) Agreement by song origin. Underlined labels in the x-axis representing local music to the rater. (C) Between country pairwise comparison using Pearson correlation corrected for attenuation. All error bars represent 95% CI.

*4.2.1 Agreement on Moods*

All three countries showed high within-country agreement for moods such as *energy*, *danceable*, *electronic*, *sad*, *cheerful*, and *love* (all $ICC_{mean} > .64$; see Figure 4A), some of which were almost comparable to the baseline measure - singer's gender ($ICC_{mean} = .94$). Certain mood variables had a stronger agreement in some countries than in others. For instance, Koreans had a lower agreement for *calm* (.49) compared to the other two countries (> .70).

There was a strong agreement about *dreamy* among the Americans (.73) but not within Koreans (.38) and Brazilians (.52). The agreement about *tense* was generally weakest across all three countries (US: .57, S. Korea: .37; Brazil: .38). Across all mood variables, ratings were more consistent among the raters in Brazil ($M = .73$, $SD = .16$) and the US ($M = .74$, $SD = .10$) than in S. Korea ($M = .63$, $SD = .17$).

*4.2.2 Agreement on Local vs. Foreign Music*

We also examined the raters' agreement for their local music versus foreign music by separating songs according to their origin and averaging across the mood variables (see Figure 4B). Koreans (orange bars) generally showed only a moderate amount of agreement for Brazilian and American music, but a substantially higher agreement for their local music ($F(2,321) = 20.7$, $p < .001$, $ges = .114$). Meanwhile, Brazilians and Americans did not particularly converge more for their local music over foreign music.

We found that Korean music received strong agreements from all three countries, with no observable group differences ($F(2, 321) = 0.53$, $p > .05$, $ges = .003$). Comparably, Brazilian music and American music received the highest agreement from their local raters but with no significant differences (Tukey's HSD with adjusted $p$s > .05).

The observed pattern is interesting, as one might expect that the intended emotional cues are most easily recognized within the same culture and therefore better converge on their agreement. Previous works have supported this [12,13], contrary to the results we observed. Moreover, while Korean songs were unfamiliar to Brazilians and Americans, they showed high convergence, suggesting that there may be distinct or *cliché* acoustic properties in Korean music that are easily recognized and agreed upon across different cultures.

**4.3 Between Country Agreement**

We next investigated how the mood perception in different dimensions aligns between the countries (i.e., BR vs. US, US vs. KR, KR vs. BR) by aggregating rater responses in each country across the song items. When computing correlations between the rater groups, the correlation is attenuated due to the measurement errors in each group. Thus, we corrected for this attenuation [33] by accounting for the internal reliability, which we estimate by the mean ICC values we computed in section 4.2. All correlation hereafter is reported using this corrected version, but the raw correlation is also reported in supplementary S3. In addition, confidence intervals were obtained by sampling 1,000 bootstrapped values with replacement.

When comparing the alignments between countries, we found that four variables, *danceable, energy, electronic,* and *sad*, are very strongly correlated among all the between-country pairs (all $r$s > .93; see Figure 4C). These four variables were also features that are directly comparable with Spotify's mood detection algorithms. In contrast, considerably lower correlations were observed between Brazilians and Koreans for *love* ($r = .63$) and *dreamy* ($r = .41$). Interestingly, these two countries exhibited much stronger correlations when paired with the US raters (all $r$s > .83). The ratings were highly similar between Brazilian and American raters for *tense,* but showed considerably weaker agreement between the Brazilian vs. Korean and American vs. Korean pairs, possibly due to the shortage of excerpts in our stimuli set manifesting this description.

Together, these results suggest that more basic mood attributes may be perceived similarly across the cultures, while more complex attributes may be perceived differently depending on the listener's cultural background.

**4.4 Human vs. MIR**

How well does MIR approximate human judgment? Is there a cultural bias in the algorithm? To examine these set of questions (RQ3), we computed the correlation between all human raters (regardless of nationality) with MIR estimations. We found that on average those correlations range between 0.49 and 0.63 across the variables *energy*, *danceable, sad,* and *electronic* (see Figure 5A; all *p*s < .001). These values were substantially lower than the mean correlations observed between the human raters for the same four features (all *r*s > .93).

Although the alignment of perceived mood ratings with Spotify's features is not weak, such a clear congruence across different cultures suggests that mood detection algorithms can be improved further to better capture these high-level perceptions that are cross-culturally consistent.

*4.4.1 Is there a Cultural Bias?*

Next, we investigated whether MIR is biased to a particular cultural group. Figure 5B shows the average correlations between human ratings and MIR values across individual song-rate pairs in each country. We found no significant differences between any of the country-country pairs using bootstrapping (*p*s > .20) and also by using a method [34] for significance testing between correlations (*p*s > .18). Thus, we conclude that there is no clearly observable bias in MIR in favoring a certain culture, contrary to our initial prediction that it may align better with the US raters.

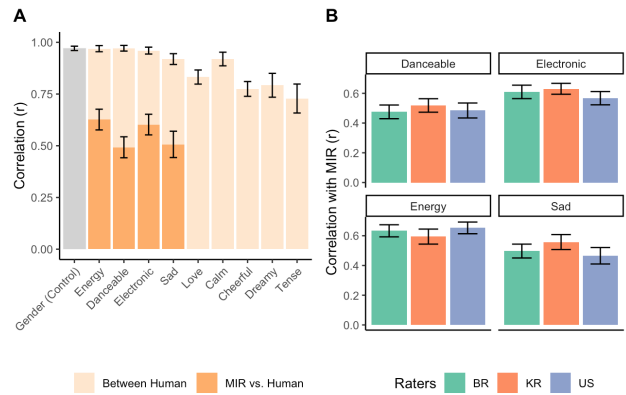

**Figure 5.** (A) Correlations among people are presented in light orange while correlations between people and MIR are in dark orange. The grey bar represents the baseline measure. (B) Correlations between MIR and raters from three countries. Error bars represent 1SD and all correlations are corrected for the attenuation.

*4.4.2 Song Familiarity and MIR*

Considering that our participants had a wide range of degrees of familiarity with songs, we also tested whether prior experience with a song influenced the alignment with MIR. We divided the songs into familiar and unfamiliar categories and found that all four MIR features were more strongly correlated with unfamiliar songs than familiar songs: *danceable* ($M_{diff}$ = 0.13, $p$ = .16), *electronic* ($M_{diff}$ = 0.16, $p$ = .02), *energy* ($M_{diff}$ = 0.27, $p$ < .001), and *sad* ($M_{diff}$ = 0.10, $p$ = .38), with p-values corrected using Bonferroni correction. This result is consistent with the idea that the listener's personal relationships to music may result in higher variability in mood perception (e.g., due to autobiographical memories [35]), which may have been reflected as reduced correlations with the objective algorithms.

## 5. DISCUSSION

We compared participant ratings from Brazil, South Korea, and the US and examined within and between-country differences in listeners' perception of nine mood attributes in music. We also assessed the robustness and generalizability of automatic mood detection algorithms in MIR by comparing it with human raters from those three countries.

Our findings reveal that relatively simple mood attributes such as *danceable*, *energy, sad, cheerful,* and *electronic* are highly agreed upon among the listeners both within and across cultures. This result suggests that pop songs may have intrinsically similar acoustic properties that are reliably recognized regardless of the listener's cultural background, even when they are unfamiliar with the musical style [7,8]. Some of these properties are low-level features (e.g., tempo, loudness) that can also be picked up by a mood detection algorithm as we found the algorithms to be quite good at approximating human judgments. Nonetheless, the substantial gap between the correlations among humans and correlations of human vs. MIR reveals the current limitation of algorithms failing to capture the full extent of human perception. This implies that people's mood perception in music may not be explained solely by the acoustic properties. However, recent advances in mood classification incorporating mid-level features [36] (e.g., melodiousness) and multi-modalities [37,38] (e.g., fusing audio and lyrics) show promising paths for improvement.

Our results also show that complex mood attributes may not have concretely shared musical characteristics and are thus perceived relatively differently depending on the listener's cultural background [12-15]. For instance, the ratings for mood *dreamy* converged well only among the Americans but not within and between the Korean and Brazilian raters. The mood *love* was strongly agreed upon within all three countries, but there was little agreement between Koreans and Brazilians. These cross-cultural differences are striking considering both South Korea and Brazil are highly globalized and thus largely exposed to the mainstream media. We would expect significantly larger cross-cultural variation in small-scale societies [39,40] and future research can look to incorporate more diverse musical styles and participant groups.

Automatic mood detection algorithms in MIR are generally trained on English songs, with the ratings obtained from Western annotators. Given this circumstance, we predicted that the algorithm would be culturally biased and align better with the US raters. However, counter to our intuition, we found no cultural bias in the algorithms. All four mood features in our comparison (*danceable, energy, sad,* and *electronic*) aligned similarly well with raters from the three countries. Given these outcomes, we conclude that current mood detection algorithms are good objective proxies for human judgments and are not culturally biased, at least within the popular music context in industrialized societies.

## 6. DATA AVAILABILITY

Raw participant ratings, musical stimuli used in the experiment, and additional statistical results are available at https://osf.io/3uw9d/

## 7. ACKNOWLEDGEMENT

We greatly appreciate the support we received from Fernanda Fernandes and Seunghyun Jeong for the recruitment of participants in Brazil and South Korea. We would also like to thank the four anonymous reviewers who provided constructive feedback. This work was funded by the MPI for Human Cognitive and Brain Science (studentship awarded to HL) and MPI for Empirical Aesthetics (NJ).